\documentstyle[epsf]{article}
\normalsize

\newcommand{\bphi}{\mbox{\boldmath $\phi$}}

\newcommand{\bPhi}{\mbox{\boldmath $\Phi$}}

\newcommand{\half}{\frac{1}{2}}
\newcommand{\pl}{\,\,\, + \,\,\,}
\newcommand{\mi}{\,\,\, - \,\,\,}

\newcommand{\kk}{{\bf k}}
\newcommand{\jj}{{\bf J}}

\newcommand{\GG}{{\bf G}}
\newcommand{\vv}{{\bf v}}
\newcommand{\uu}{{\bf u}}

\newcommand{\dg}{\dagger}
\newcommand{\pd}{\partial}
\newcommand{\ba}{\begin{eqnarray*}}
\newcommand{\ea}{\end{eqnarray*}}
\newcommand{\be}{\begin{equation}}
\newcommand{\ee}{\end{equation}}
 
\begin{document}
 
\bibliographystyle{unsrt}
\footskip 1.0cm
\thispagestyle{empty}
\setcounter{page}{0}
\begin{flushright}
Imperial/TP/96-97/14\\
Heidelberg/HD-THEP-96-59\\
December 1996\\
\end{flushright}
\vspace{10mm}
 
\centerline {\LARGE Non-Abelian String Conductivity}
\vspace*{15mm}
\centerline {\large T.W.B. Kibble$^{(1)}$, G. Lozano$^{(2)}$ and
A.J. Yates$^{(1)}$ } 
\vspace{5mm}
\centerline {\it {}$^{(1)}$Imperial College, Theoretical Physics}
\centerline{\it Prince Consort Road, SW7 2BZ London, England}
\vspace*{3mm}
\centerline {\it {}$^{(2)}$Universitat Heidelberg, Institut fur
Theoretische Physik} 
\centerline{\it Philosophenweg 16, D69120 Germany} 
 
\vspace{12mm}
\normalsize
\oddsidemargin 0.3cm
\evensidemargin 0.3cm
\centerline{\bf Abstract}
\vspace{8mm}
We examine current-carrying configurations of cosmic strings in non-Abelian
gauge theories. We study the solutions numerically
and point out that the currents will be at best {\it
dynamically}\, stable and not subject to any topological quantisation or
conservation, as in conventional models of string
superconduction. We suggest that non-Abelian string
loops may be unable to support persistent currents in the absence of
external fields. This will have relevance to vorton stability.

\setcounter{page}{1}
\oddsidemargin 0.13cm
\evensidemargin 0.13cm
\section{Introduction}
It has been known for several years that many cosmic string models
appear to exhibit superconductivity~\cite{Witten}. The currents may be
very large for strings formed during the Grand-Unified phase
transition, and may have significant cosmological implications.
For instance, it has been argued that decaying superconducting loops
can provide an explosive  mechanism for galaxy formation
\cite{Ostriker} (although 
it is not clear if this scenario is compatible with the observed
anisotropies in the microwave background \cite{Levin}). Observational
bounds on cosmic rays from superconducting strings networks \cite{Hill} and
the synchroton radiation signatures of these  topological
defects have been also discussed \cite{Chud}. String loops stabilised
against collapse by the effect of a current flowing on them -- {\it
vortons} \cite{vortons} -- are a potential dark matter candidate, and
indeed if they 
exist from the time of the  GUT scale they would present an
overdensity problem that could rule out string scenarios \cite{Everyone}.

In his original paper \cite{Witten}, Witten proposed two mechanisms
for cosmic string superconductivity. The first is based on
a charged {\it scalar boson} field which condenses in the core
of the string, breaking electromagnetic gauge invariance there and
supporting supercurrents in a manner directly analogous to
conventional BCS theory. The second involves charged
{\it fermionic} zero modes trapped in the core of the string and travelling 
along it at the speed of light. Both mechanisms are shown
to operate in a simple U(1)$\times$U(1) model,  although for the
scalar-bosonic case the presence of superconductivity requires a rather
intricate fine-tuning of parameters -- indicating that this is
unlikely to be a generic feature of string-forming models.
Despite this, as the model is sufficiently simple it has been studied
thoroughly and the astrophysical consequences of superconducting
strings have been explored by many authors.
We refer the reader to the reviews \cite{reviews} for more details.

A distinct and more natural mechanism for string conductivity arises
in non-Abelian string models, that is free of the coupling dependence
that restricts the mechanisms above. Here the
charge carriers are (super-heavy) vector bosons which, if absent in the model
mentioned above, are present in any Grand-Unified model. This
possibility was first mentioned by Preskill \cite{Presk} and it was
subsequently considered 
to some extent by Everett \cite{Everett} and Alford {\it et al.}
\cite{Alford}.
By vector-boson conductivity we are {\it not} referring to the
phenomenon of W-condensation near an ordinary GUT cosmic string, which
could be the source of electromagnetic currents at the electroweak
scale (see, for instance, references \cite{Amb}-\cite{AnneWarren2}),
but rather the case in which 
the string itself is constructed from electromagnetically charged
vector bosonic fields. 

In this paper we will attempt a more systematic study of the
conduction properties of this type of string. In section
\ref{section:Witten} we will briefly  review the basics of scalar
boson superconductivity in the U(1)$\times$ U(1) 
model, analysing the cases of a straight infinite string and a
string loop. This provides the motivation for some distinctions made
later in the paper. In section \ref{section:simple_z2_string} we
introduce a toy model (SU(2)$\times$U(1) with a complex Higgs in the
adjoint representation) which is the simplest theory that supports
conducting non-Abelian strings, and 
allows us to study their electromagnetic properies
without going through the technical complications of examining a realistic
Grand-Unified group. We show in section
\ref{section:SO_10}, however, that this model 
can be embedded directly in SO(10) with the Higgs in the {\bf 126}
representation, which is the simplest Grand-Unified theory exhibiting
topologically stable strings. Finally, in section
\ref{section:conclusion} we explain why the electrodynamic properties
of loops of non-Abelian conducting string appear to
be very different to those of their Abelian counterparts, 
speculate about the cosmological implications and summarize our results.

\section{Scalar boson superconductivity}
\label{section:Witten}
Superconductivity can be understood generally as broken
electromagnetic gauge invariance.  In general this appears when the
expectation value of a charged, 
bosonic condensate or order parameter is non-zero.  Witten
\cite{Witten} examined a U(1)$\times$U(1) model in which a charged
scalar $\phi$ condensed around the core of an uncharged Abelian
string, breaking electromagnetic symmetry 
there and allowing currents to flow along the string.  We review it
briefly here in order to introduce some concepts that will be useful later.

We begin with a Lagrangian with U(1)$\times$U(1) symmetry:
\[
{\cal L} = - (D_\mu \Phi)^* D^\mu \Phi - (D_\mu \phi)^* D^\mu \phi
- \frac{1}{4}G_{\mu\nu} G^{\mu \nu}
- \frac{1}{4}F_{\mu\nu} F^{\mu \nu} - V(\Phi,\phi),
\]
where we use a polar coordinate metric\footnote{Throughout the paper we will
use  convention in which indices $\mu ,\nu,..$ run from 0 to 3;
$\alpha, \beta,...$ denote $z,t$ and $i,j,..$ denote $r,\theta$.}
$g_{00}=-1$, $g_{zz}=g_{rr}=1$, 
$g_{\theta \theta}= r^2$ and 
\ba
 D_\mu\Phi &=&(\partial_\mu - igR_\mu)\Phi \\
 D_\mu \phi &=&(\partial_\mu - ieA_\mu)\phi \\
 G_{\mu\nu} &=& \partial_\mu R_{\nu} - \partial_\nu R_{\mu} \\
 F_{\mu\nu} &=& \partial_\mu A_{\nu} - \partial_\nu A_{\mu} \\
V(\Phi,\phi) & = &\half \alpha(|\Phi|^2 - \sigma^2)^2 +
\half \chi(|\phi|^2 - \eta^2)^2 +
\half \gamma |\Phi|^2|\phi|^2.
\ea
The U(1) symmetry gauged by $R_\mu$ is broken by the complex scalar
$\Phi$  if $\chi \eta^4 < \alpha \sigma^4$ while the U(1) gauged
by $A_\mu$ remains unbroken if  $\gamma \sigma^2 /2 >
\chi \eta^2$. Nevertheless, in the presence of a string 
 it may be favorable for $\phi$ to acquire a vacuum expectation value in the
core of the string (where $\Phi$ is zero), effectively breaking the
U(1) of electromagnetism there but not outside the string. This is indeed
the case when, for instance, 
\be
\gamma \sigma^2 /2 \gg \chi \eta^2 \qquad 
\left( \frac{\sigma}{\eta} \right)^3 > \left( \frac{\chi}{\alpha} \right)^2
\qquad
\frac{\gamma}{\chi} \, \stackrel{<}{\sim} \,\frac{\chi \eta^4}
{\alpha \sigma^4}.
\ee
The {\it ansatz} for the bare string, with no current, is
\be
R_\theta=b(r) \qquad \Phi(r,\theta)=f(r)e^{i\theta}  \qquad
\phi(r,\theta)=\phi(r), 
\ee
where $b(r)$ and $f(r)$ have the usual Nielsen-Olesen profiles and 
\be
\phi(0) \neq 0, \qquad \phi(\infty)=0.
\ee

Let us then examine how currents arise in this model.
Clearly the configuration above
must be extended to introduce some $z,t$ dependence of the
fields, as well as a non-zero electromagnetic field. 
The most natural way to do this is with
\begin{eqnarray}
\phi(r) &\rightarrow & \phi(r) e^{i c_0 \beta(z,t)}
\nonumber \\
A_\alpha &\rightarrow & -\frac{1}{e} s(r) \partial_\alpha \beta(z,t) \qquad
\alpha \in \{z,t\}
\label{eqn:an12} 
\end{eqnarray}
and keeping the other fields as before. The factor $c_0$ could, of
course, be absorbed into the function $\beta(z,t)$ but it is useful to
retain it, as we shall see below.
It appears initially that
we could simply remove the  $z,t$ dependence of
$\phi$ with a gauge transformation. This is obviously true for the
case of an infinite string,  but as we will see it is essential to
consider the boundary
conditions, and the situation is more delicate when we extend the 
analysis to the case of string loops.  

In addition, we will demand that $\eta(z,t)$ satisfies the transverse
field equations,
\be
\partial_\alpha \partial^\alpha \beta =0.
\ee
Introducing this {\it ansatz} into the Euler-Lagrange equations,
one finds that for consistency one also has to demand
\be
\partial_\alpha \beta \partial^\alpha \beta =\epsilon,
\ee
where $\epsilon$ is a constant. This last constraint arises as a result of
our attempt to use separation of variables in a non linear equation.
Although it might appear too restrictive, notice that it includes
three relevant physical cases,
\be
\beta=kz \qquad \beta=\omega t \qquad 
\beta=k(z-t),
\ee
corresponding to a static current, a static charge density  and
a charged current pulse travelling at the speed of light
respectively. This last case, 
($\epsilon=0$, or the so-called `chiral' state) is special, 
since the backreaction of the electromagnetic fields on
string vanishes. We will return
to this point in studying the non-Abelian model. 

The equations of motion for the $z,t$ components of the gauge fields
reduce to
\begin{equation}
\label{eqn:s_eqn}
\nabla_r^2 s(r) =  2 e^2 |\phi(r)|^2 (s(r)-c_0).
\end{equation}

As we have indicated, in the case of a straight infinite string
the constant $c_0$ is irrelevant since the phase of the condensate can
be removed  with a gauge transformation. In this case, then, 
we have a {\it homogeneous} equation for $s(r)$
\begin{equation}
\label{eqn:s_eqn2}
\nabla_r^2 s(r) =  2 e^2 |\phi(r)|^2 s(r).
\end{equation}
This equation certainly admits the trivial solution $s(r)=0$,
unless of course  we impose a non trivial
boundary condition at infinity -- specifically,
\be
s'(r) \rightarrow -\frac{I e }{2 \pi r},
\ee
representing the presence of a current (or charge, or both)
on the string. The key to the presence of supercurrents
is the condensate $|\phi(r)|$ which is non-vanishing at the origin 
(the core of the string), and without which we would not have 
a regular solution to equation (\ref{eqn:s_eqn}).

The case of a string loop is significantly different, 
however. We restrict ourselves for the moment to the uncharged
case $\eta=kz$, where now we designate with $z$ the coordinate
along the loop (which we take of a large radius $R$ in order
to avoid curvature effects). Naturally, in order for the condensate
field to be single-valued, we require that $c_0 2 \pi R k =N$ where
$N$ is an integer. Now the phase of $\phi$ has a clear physical meaning --
it {\it cannot} be removed by a gauge transformation and is something
that the loop can acquire on becoming superconducting.
In another words, if one assumes (following Davis and Shellard
\cite{DavisShell})
that an ansatz of the type given by equation (\ref{eqn:an12}) will
still be valid on a 
loop, the constant $c_0$ in equation (\ref{eqn:s_eqn}) being nozero,
means that the gauge fields are {\it sourced} by the
phase winding number. Notice that the role played
for the straight string by the boundary condition at infinity
translates on a loop into the specification of $N$, which is related
to the total current.

In the following section we will confine ourselves to
the infinite straight string case, and study the generalisation
of equation (\ref{eqn:s_eqn}) for both a simple non-Abelian model and
the well-known SO(10) string. Note that a crucial distinction arises
when we discuss the case of loops, however, which we mention here. The
extension  
of the comments above regarding the topologically conserved phase of
the scalar condensate 
does not appear to be possible for a general non-Abelian theory -- in
fact, it appears to be impossible to construct field configurations
with any 
such topological invariants -- and it is more appropriate to view the
currents as carried purely by gauge field components. This is a
distinction analogous to the difference between {\it super}conductors, which
carry currents non-dissipatively and, in addition,  support conserved
currents in multiply-connected samples in the absence of external
fields, and {\it perfect} conductors, which display similar bulk features
(the Meissner effect, for instance) but whose currents appear and
disappear reversibly in the presence or absence of an external field,
with no possibility of `topological' preservation.
This  may have
important cosmological consequences -- 
particularly for studies of vorton stability. We return to this point
in the conclusion.

\section{A non-Abelian string model: SU(2)$\times$U(1)}
\label{section:simple_z2_string}
\subsection{The model}

To investigate non-Abelian string conductivity, we
consider a model which will bring out the
relevant features of the mechanism. In the next section we look at its
embedding in more realistic GUT  model.

The simplest non-Abelian model that displays string solutions is given
by the breaking of SU(2) to $Z_2$ by two real Higgs in the adjoint
representation. In this case SU(2) is effectively SO(3). The homotopy
group $\pi_1({\rm SO(3)}/1)=Z_2$ and so topologically stable strings are
formed. Different {\it ans\"atze} leading to string solutions
have been analysed in the past (see for instance \cite{deVega}, 
\cite{Khare}) 

Since all the symmetries are broken in this
case, to provide for the possibility of a long-range massless field
playing 
the role of electromagnetism we extend the theory  to
SU(2)$\times$U(1). We do this by arranging the two real Higgs
in a complex multiplet and gauging the corresponding U(1) symmetry.

The model is then described by the following Lagrangian density:
\be
{\cal L} = -\frac{1}{4} F_{\mu \nu}F^{\mu \nu} - \frac{1}{4}{\bf
G}_{\mu \nu} \cdot {\bf G}^{\mu \nu} - D_\mu \bphi^\dg\cdot D^\mu {\bf
\bphi} - V(\bphi), 
\ee
where 
\begin{eqnarray}
\label{eqn:definitions}
F_{\mu \nu} & = & \partial_\mu A_\nu - \partial_\nu A_\mu, \nonumber \\
{\bf G}_{\mu \nu} & = & \partial_\mu {\bf B}_\nu - \partial_\nu {\bf
B}_\mu + g {\bf B}_\mu \times {\bf B}_\nu, \nonumber \\
D_\mu \bphi & = & \partial_\mu \bphi - ie A_\mu \bphi +
g {\bf B}_\mu  \times \bphi, \nonumber \\
V(\bphi) & = & \half \lambda(\bphi^{\dg}\cdot \bphi - \half
\eta^2)^2 + \half \kappa |\bphi \cdot \bphi|^2.
\end{eqnarray}
Here, ${\bf B}_\mu$ and $\bphi$ are three-dimensional vectors in the
SU(2) Lie Algebra and  ``$ \cdot$" and ``$\times$" denote the standard
scalar and cross product in 
internal space.

If we choose $\kappa> 0$, then the vacuum is characterised by
\be
\bphi \cdot \bphi = 0, \qquad \bphi^\dg \cdot \bphi
= \half \eta^2.
\ee
We then define the vacuum to be
\be
\bPhi_0={\eta\over 2}\pmatrix{1\cr 0\cr i}.
\ee
Let the SU(2) generators  in the adjoint representation 
be $T_i$ ( $i=1,2,3$) and the U(1) generator be $T_0$,
\be
(-iT_i \bphi)_j = -\epsilon_{ijk}\bphi_k, \qquad (-iT_0 \bphi)_j = -i
\bphi_j.
\ee
The generator that annihilates $\bphi_0$, and which we identify
with electromagnetic charge, is
\be
 Q= T_2 + T_0.
\label{eqn:char1}
\ee

\subsection{Bare string solutions}
This theory supports topologically stable strings. Imagine that a
string is formed, by $\bphi$ acquiring its non-zero expectation value
above in the usual way, and take the string generator to be
\be
T_S= T_3.
\ee
At large distances from the string core, 
\be
\bphi(\theta)
 = e^{-i\theta T_3} \bPhi_0 = \frac{\eta}{2} \pmatrix{\cos \theta
\cr \sin \theta \cr i}.
\label{eqn:sca1}
\ee
A distinctive feature of our ansatz is that the
string generator does not commute with the charge generator (given
by equation (\ref{eqn:char1})):
\be
[T_S, Q]=[T_3,T_2]= -i T_1.
\label{eqn:nonc}
\ee
As a result, in the presence of this string the generator $Q$
which annihilates the vacuum acquires an angular dependence:
\be
 Q(\theta)= e^{- i \theta T_S} Q e^{ i \theta T_S}= T_0 + T_2 \cos{\theta}
- T_1 \sin{\theta}.
\ee
We define the following unit vectors in internal space:
\be
{\bf e}_1 = \pmatrix{\cos \theta \cr \sin \theta \cr 0}, \qquad 
{\bf e}_2 = \pmatrix{-\sin \theta \cr \cos \theta \cr  0}, \qquad
{\bf e}_3 = \pmatrix{0 \cr 0 \cr 1}.
\label{eqn:base1}
\ee
The {\it ansatz} for the scalar field is then
\be
\bphi(r, \theta)
 = \frac{\eta}{2} \left({\bf e}_1 f(r) + i {\bf e}_3 h(r)\right).
\label{eqn:ans}
\ee
The real functions $f$ and $h$ satisfy
\be
f(0)=0, \qquad f(\infty) = h(\infty) = 1,
\ee
and there is no topological restriction on $h(0)$. We shall see below
that it can be energetically favourable (in the absence of a current)
for this component to persist to the core of the string, to reduce
both gradient and potential energy.

Although the basis given by equation (\ref{eqn:base1}) is the most
convenient for the 
calculation, we will introduce two other bases which will 
be useful for the physical interpretation of the solutions.
One of them is given by the eigenvectors of the string generator,
\be
\uu_{+1} = \frac{1}{\sqrt{2}}\pmatrix{1 \cr i \cr 0}, \qquad 
\uu_{-1} = \frac{1}{\sqrt{2}}\pmatrix{1 \cr -i \cr  0}, \qquad
\uu_0 = \pmatrix{0 \cr 0 \cr 1},
\label{eqn:base2}
\ee
and the other by the eigenvectors of $Q(\theta)$;
\be
\vv_0= \frac{1}{\sqrt 2} ({\bf e}_1 + i {\bf e}_3), \qquad
\vv_{+1}= {\bf e}_2, \qquad
\vv_{+2}= \frac{1}{\sqrt 2} ({\bf e}_1 - i {\bf e}_3).
\label{eqn:base3}
\ee
In term of these bases, the scalar field is expressed as
\begin{eqnarray}
\label{eqn:base_decomp}
\bphi(r,\theta)  &=&  \frac{\eta}{2} \left( \frac{f(r)}{\surd 2}
(e^{i\theta} \uu_{+1} +  e^{-i \theta} \uu_{-1}) + i h(r) \uu_0 \right),
\nonumber \\
 &=&  \frac{\eta}{2} \left( \frac{f(r) + h(r)}{\surd 2} 
\vv_0 +  \frac{ f(r) - h(r)}{\surd 2} \vv_{+2}  \right).
\end{eqnarray}
It is interesting to point out that instead of the form
in equation (\ref{eqn:ans}), we could have chosen another {\it ansatz} for the
scalar field:
\be
\bphi_{NO}(\theta) = \frac{\eta}{2}f_{NO}(r) e^{-i\theta} \bPhi_0.
\label{eqn:ansab}
\ee
Here the string generator is proportional to 
\be
T_S^{'}= T_2-T_0,
\ee
and so
\be
[T_S^{'}, Q]= 0.
\ee
Notice also that in constrast to the form (\ref{eqn:ans}) there is only one
function $f(r)$ involved, which is constrained to vanish
at the origin since every component of the Higgs is wound by the
string generator $T_S^{'}$. In this
case, the string is just an embedding of the U(1) Nielsen-Olesen
vortex in a larger group. Which of the two {\it ans\"atze} is realised
in a given model is matter of energetics and, as we shall discuss
below,  depends largely on the parameters of the model. Indeed, in
the out-of-equilibrium conditions of a phase transition it seems
likely that all possible string configurations will arise. However,
only those in  which the string fields are charged -- $[T_s, Q]\neq 0$
-- support currents in the sense we discuss in this paper.

Returning our attention to the lower-energy solution, the {\it ansatz}
for the string gauge field is 
\be
{\bf B}_\theta =  b(r) {\bf e}_3 , 
\ee
where $b(0)=0$ and $b(\infty)= -1/g$. 
This, together with equation (\ref{eqn:ans}), completes the
{\it ansatz} for the non-conducting string fields.
The equations of motion
\ba
 \frac{1}{\sqrt{g}} \partial_\mu \sqrt{g} F^{\mu \nu} &=& j^{\alpha} =
 i e [ \bphi^{\dg} \cdot D^{\nu}\bphi - \bphi \cdot (D^{\nu} \bphi)^{\dg} ],
\\
 \frac{1}{\sqrt{g}} D_\mu \sqrt{g} {\bf G}^{\mu \nu} &=& {\bf J}^{\alpha} =
 g [ \bphi^{\dg} \times D^{\nu}\bphi + \bphi \times (D^{\nu} \bphi)^{\dg} ],
\\
 \frac{1}{\sqrt{g}} D_\mu \sqrt{g} D^\mu \bphi &=& \frac{\delta V}{\delta 
\bphi^{\dg}},
\ea
then become
\begin{eqnarray}
\label{eqn:EOM1}
\left( \frac{1}{r}b'\right)' & = & \frac{\eta^2}{2r} g (1+gb)f^2,
\nonumber \\
\frac{1}{r}(rf')' & = & \frac{1}{r^2}(1+gb)^2 f +
\frac{1}{4}\lambda \eta^2 f (f^2 + h^2 -2) + \frac{1}{4}\kappa\eta^2
f(f^2 - h^2), \nonumber\\
\frac{1}{r}(rh')' & = & \frac{1}{4}\lambda \eta^2 h (f^2 + h^2
-2)+ \frac{1}{4} \kappa\eta^2 h(h^2 - f^2).
\end{eqnarray}
The energy per unit length of this configuration is given by
\[
H_{na}  =  \int r \, dr \, d\theta \,  \left\{ \frac{1}{2r^2} b'^2 + 
\frac{\eta^2}{4}\left[f'^2 + h'^2\right] +
\frac{\eta^2}{4r^2}(1+gb)^2 f^2 + \right.
\]
\be
  \left.\frac{\lambda\eta^4}{32} \left(f^2 + h^2 - 2\right)^2 
+ \frac{\kappa\eta^4}{32}\left(f^2 - h^2\right)^2\right\}.
\label{eqn:na-ene}
\ee
We note that for $\kappa=\lambda$, the Higgs component $h$
decouples from the equations and can be set to unity everywhere.
In this limit, the equations reduce to those
of the Abelian Higgs model, whose solutions are well-known.
Further, if we now impose the additional constraint $g^2=2\lambda$,
the second order differential equations can be shown
to be equivalent to the first order Bogomol'nyi equations \cite{Bogo},
\ba
b' & = & \frac{g \eta^2}{4}r(f^2 -1) \\
f' & = & \frac{f(1+gb)}{r}.
\ea
For the Abelian {\it ansatz} of Eq.(\ref{eqn:ansab}) we take instead
\be
\alpha Z_\theta (r) = e A_\theta (r) - g B^2_\theta(r),
\ee
where $\alpha = \sqrt{e^2 + g^2}$. In this case, the equations become
\ba
\left( \frac{1}{r} Z'\right)' & = & \frac{\eta^2}{r} \alpha (1+\alpha Z)f^2, \\
\frac{1}{r}(rf')' & = & \frac{1}{r^2}(1+ \alpha Z)^2 f +
\frac{1}{2}\lambda \eta^2 f (f^2  -1),  \\
\ea
with a resulting energy per unit length
\be
H_{ab}  = \int r \, dr \, d\theta \, \left\{ \frac{1}{2r^2} Z'^2 + 
\frac{\eta^2}{2} f'^2  +
\frac{\eta^2}{2r^2}(1 + \alpha Z)^2 f^2 + 
\frac{\lambda\eta^4}{8} \left(f^2  - 1\right)^2 \right\}.
\label{eqn:a-ene}
\ee
Similar to the non-Abelian ansatz, the equations of motion also
reduce to a Bogomol'nyi form when $\lambda=\alpha^2/2$.

Notice that for the non-Abelian ansatz, $h(r)$ does not vanish
at the origin.
We can estimate its the value by equating to zero
the right hand side of the last equation, giving
\[
h(0)^2 \simeq \frac{2\lambda+ {\cal O}(h(0)''/h(0))}{\lambda
+ \kappa},
\]
which means that in the general case (that is,  unless $\kappa$ is
infinitely large) 
the full symmetry {\it is not} restored inside the string. 

\begin{figure}
\begin{center}
    \leavevmode
        {\hbox %
{\epsfxsize = 15cm\epsfysize=10cm
    \epsffile {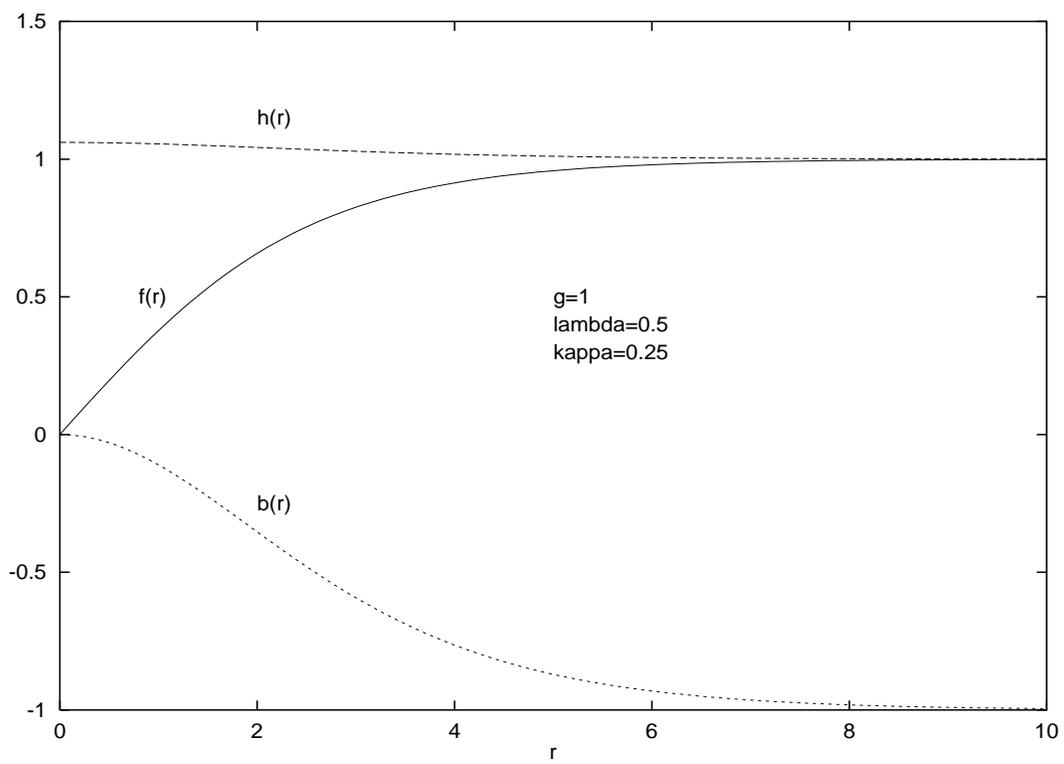} }}
\end{center}
\caption{Fields for a string formed in the breaking SU(2)$\times$U(1)
$\rightarrow$ U(1). Here $\eta = 1$.}
\label{fig:z2fields}
\end{figure}

The equations (\ref{eqn:EOM1})
can be solved  numerically, either with a
fifth-order Runge-Kutta shooting method -- giving initial estimates for
$h(0)$, $f'(0)$ 
and $b''(0)$ --  or by a relaxation method, giving initial estimates for
$h(r)$, $f(r)$ and $b(r)$. The resulting 
fields for a representative set of values for the couplings are shown
in figure \ref{fig:z2fields} (produced with the shooting method).

As we mentioned in the introduction, we wish to
embed this solution in the
SO(10) model and this will be possible provided that
a certain relation between the gauge couplings holds.  As will 
become clear in section \ref{section:SO_10}, or just
by a direct  comparison with the equations of motion in \cite{Ma}, 
the embedding of the Abelian ansatz for the SU(2)$\times$U(1) model
in SO(10) requires $g^2 = \frac{2}{3} e^2$.
In this case, it was shown by Aryal and Everett \cite{AE} and Ma
\cite{Ma} that the non-Abelian string
form has a lower energy than the Abelian one for all
parameters in the potential. This result is somewhat surprising.
Notice, for instance, that the magnetic energy  -- represented by
the first term in each of the equations (\ref{eqn:na-ene}) and
(\ref{eqn:a-ene}) --  is 
proportional to $1/g^2$ and $2/5g^2$ for the non-Abelian
and Abelian string respectively. Neither does the difference between
the two arise from the potential
energy of the Higgs, since 
for instance in the limit of very large $\kappa$, $h \rightarrow f$
in the non-Abelian ansatz, 
and both forms have the same potential energy. The reason
the non-Abelian ansatz is energetically preferred to
the Abelian one seems to arise instead from the Higgs-gauge field
interaction; while in the first case only one component of the
Higgs field ($f$) interacts with the gauge fields, in the second case
both components (which are identical) do. According to the
analyses performed in references \cite{Ma,AE}, then,
the non-Abelian string 
is energetically favoured in SO(10). We speculate that in the
out-of-equilibrium conditions of a phase transition, however, all
possible string solutions will arise -- nevertheless,
the measure of Abelian solutions is outweighed by that of the
non-Abelian form. We will show in the next section that this latter
type of string supports currents.

We mention here that a less direct mechanism by which {\it Abelian} 
strings can  become superconducting, through electroweak gauge boson
condensation, has been analysed by several
authors \cite{Amb,AnneWarren1,AnneWarren2}. 

\subsection{Incorporating a current}
Now we add fields to our solution representing a current along the
string. We allow the time and $z$-components of the gauge 
fields to be non-zero:
\ba
A_\alpha & = & a(r, \theta) \partial_\alpha \beta(z,t) \\
{\bf B}_\alpha & = & \kk (r, \theta) \partial_\alpha \beta(z,t),
\ea
where $\alpha=t,z$ and $\beta$ satisfies the transverse wave equation
\be
\partial_\alpha \partial^\alpha \beta = 0,
\ee
The equations of motion are now (where $i,j=r$,$\theta$)
\ba
\frac{1}{\sqrt{g}} \partial_i \sqrt{g} \partial^i A^\alpha &=& j^\alpha \\
\frac{1}{\sqrt{g}} D_i \sqrt{g} D^i {\bf B}^\alpha &=& {\jj}^\alpha
\ea
While the equation of motion for the transverse components of $A$ 
is not modified,
\be
\frac{1}{\sqrt{g}} \partial_i \sqrt{g} F^{ij} = j^j,
\ee
there are new terms in the equations for ${\bf B}_i$ and $\bphi$ which
represent the backreaction of the currents on the string fields:
\be
\frac{1}{\sqrt{g}} D_i \sqrt{g} \GG^{ij} = \jj^j + \jj^j_{BR}
\ee
\be
\frac{1}{\sqrt{g}} D_i \sqrt{g} D^i \bphi = 
\frac{\delta V}{\delta \bphi^{\dg}} + F_{BR}
\ee
where
\be
\jj_{BR}^{j}= g \kk \times D^j \kk \partial_\alpha \beta
\partial^{\alpha}\beta, 
\ee
and
\be
F_{BR} = [( -e^2 a^2 \bphi -2 ieg a \, \kk \times \bphi
+ g^2 \kk \times (\kk \times \bphi) ] \, \partial_\alpha \beta
\partial^{\alpha} \beta.
\ee
We see here that the backreaction vanishes in the so-called `chiral'
state -- that is, when $\beta$ is function of light-cone coordinates
$\beta=\beta(z\pm t)$.

Taking the form for the string background $\bphi$ in equation
(\ref{eqn:ans}), and 
\be
\kk = k(r) {\bf e}_2
\ee
(all other components of $\kk$ decouple and can be set to zero),
the current field equations are
\begin{eqnarray}
\label{eqn:curr_EOM}
(ra')' & = & \half \eta^2 r [ e^2(f^2 + h^2)a - 2egfhk]\nonumber \\
(rk')' & = & \frac{1}{r} (1+gb)^2 k + \half \eta^2 r [g^2 (h^2 + f^2)k
-2egfha] 
\end{eqnarray}
We see that for consistency in the transverse field equations, 
we have to demand
\be
\partial_\alpha \beta \partial^\alpha \beta = \epsilon.
\ee
where $\epsilon$ is an arbitrary constant, so that all transverse
field components are independent of $z$ and $t$.  We see then that the
string equations  of motion are modified in the following way:
\begin{eqnarray}
\label{eqn:str_EOM}
\left( \frac{1}{r}b'\right)' & = & \frac{\eta^2}{2r} g (1+gb)f^2 +
\epsilon\frac{1}{r} g (1+gb)k^2, \nonumber\\
\frac{1}{r}(rf')' & = & \frac{1}{r^2}(1+gb)^2 f +
\frac{1}{4}\lambda \eta^2 f (f^2 + h^2 -2) + \frac{1}{4}\kappa\eta^2
f(f^2 - h^2) +\nonumber \\
& & \epsilon ( e^2 a^2 f - 2eg a k h + g^2 k^2 f)
,\nonumber \\
\frac{1}{r}(rh')' & = & \frac{1}{4}\lambda \eta^2 h (f^2 + h^2
-2)+ \frac{1}{4} \kappa\eta^2 h(h^2 - f^2) +\nonumber \\
& & 
\epsilon ( e^2 a^2 h - 2eg a k f + g^2 k^2 h).
\end{eqnarray}
Before entering into a full numerical study of the equations, we can see
the pertinent features of the solutions by making some
approximations. We 
can analyse the asymptotic behavior of the electromagnetic field
by using an
approximate form for the string background to solve equations
(\ref{eqn:curr_EOM}) -- namely, 
\begin{enumerate}
\item $r<R, \;\; f=b=0, \;\; h=h_0 = $ constant
\item $r>R, \;\; b=-1/g, \;\; f=h=1.$
\end{enumerate}
We have in region (1)
\ba
\frac{1}{r}(r a')' & = & \frac{\eta^2 e^2 h_0^2}{2} a \\
\frac{1}{r}(r k')' & = & \left(\frac{1}{r^2} +
\frac{\eta^2 g^2 h_0^2}{2} \right)k,
\ea
and in region (2)
\ba
\frac{1}{r}(r a')' & = & e \eta^2 (ea-gk),\\
\frac{1}{r}(r k')' & = & g \eta^2 (ek-ea).
\ea
If we transform to our new variables
\be
\gamma = \frac{ga + ek}{\sqrt{e^2 + g^2}}, \qquad
Z = \frac{ea - gk}{\sqrt{e^2 + g^2}},
\ee
we see that
\ba
\frac{1}{r}(r Z')' & = & (e^2 + g^2) \eta^2 Z, \\
\frac{1}{r}(r \gamma')' & = & 0. 
\ea
Now we solve these equations, satisfying the the boundary conditions
we require. In region (1),
\ba 
a & = & {\cal C}_1 I_0 \left(\frac{\eta e h_0}{\surd 2}r\right), \\
k & = & {\cal C}_2 I_1 \left(\frac{\eta g h_0}{\surd 2}r\right),
\ea
where $I_0$ and $I_1$ are modified Bessel
functions that are regular at the origin, 
so that $a\sim$ constant and $k \sim r$ at $r=0$, and in
region (2)
\be
Z(r)  =  {\cal C}_3  K_0 \left(\sqrt{e^2 + g^2})\eta r\right),\\
\ee
\be
\gamma(r)  =   {\cal C}_4\ln{\frac{r}{R}} + {\cal C}_5, 
\label{eqn:gamma}
\ee
so that $Z$ decays exponentially at large $r$.
The five arbitrary constants are related by the four continuity
conditions at $r=R$, and so as expected
there remains one arbitrary constant that sets the magnitude of the
current. The effective electromagnetic current on the string
is obviously related to the asymptotic magnetic field $B$ by the
relation 
\be
B(r) = \pd_r \gamma = \frac{I}{2\pi r},
\ee
and so
\be 
I = 2\pi {\cal C}_4.
\ee

We can now study the back reaction of this current on the string by
studying the linearised version of equations (\ref{eqn:str_EOM})
in the presence of an electromagnetic gauge field of the
form (\ref{eqn:gamma}). Defining
\be
b(r)= -\frac{1}{g} + \sqrt{r} X_1(r) \qquad
f(r)-h(r)=\frac{1}{\sqrt{r}} X_2(r), 
\ee
noting that clearly only the charged linear combination $f-h$ couples
to the current, one obtains the following at large distances from the core:
\ba
& & X_1'' - \lambda_1(r) X_1 =0, \\
& & X_2'' -\lambda_2(r) X_2= 0,
\ea
where
\ba
\lambda_1(r) &=& \frac{g^2}{2} \left( 1 + \epsilon \frac{2 e^2}{e^2 + g^2}
\gamma^2(r)\right) \\ 
\lambda_2(r) &=& \kappa \pl \epsilon \frac{4 e^2 g^2}{e^2 + g^2} \gamma^2(r)
\ea
For $\epsilon > 0$, the current backreaction completely
dominates the asymptotic behavior of the fields;
\be
X_i =  \exp{\left( -|c_i| r \ln r \right)},
\ee
where the constants $c_i$ are
\be
c_1^2= \frac{e^2 g^2}{e^2+g^2} \frac{I^2}{4\pi^2}, \qquad c_2^2=4 c_1^2.
\ee
 
For the case $\epsilon < 0 $ (a string carrying a pure charge) we
cannot go to arbitrarily large distances from the core, since 
$\lambda_i(r)$ would turn negative and we would obtain oscillatory
solutions which would produce 
a linearly divergent energy per unit length. Solutions for this
case may have a physical meaning if there is a natural large-distance
cut-off in the problem, as can be for the case of loops (a situation
analogous to this occurs for Abelian strings -- see, for instance, the
study by Davis and Shellard \cite{DavisShell}). 

\begin{figure}
\begin{center}
    \leavevmode
        {\hbox %
{\epsfxsize = 15cm\epsfysize=10cm
    \epsffile {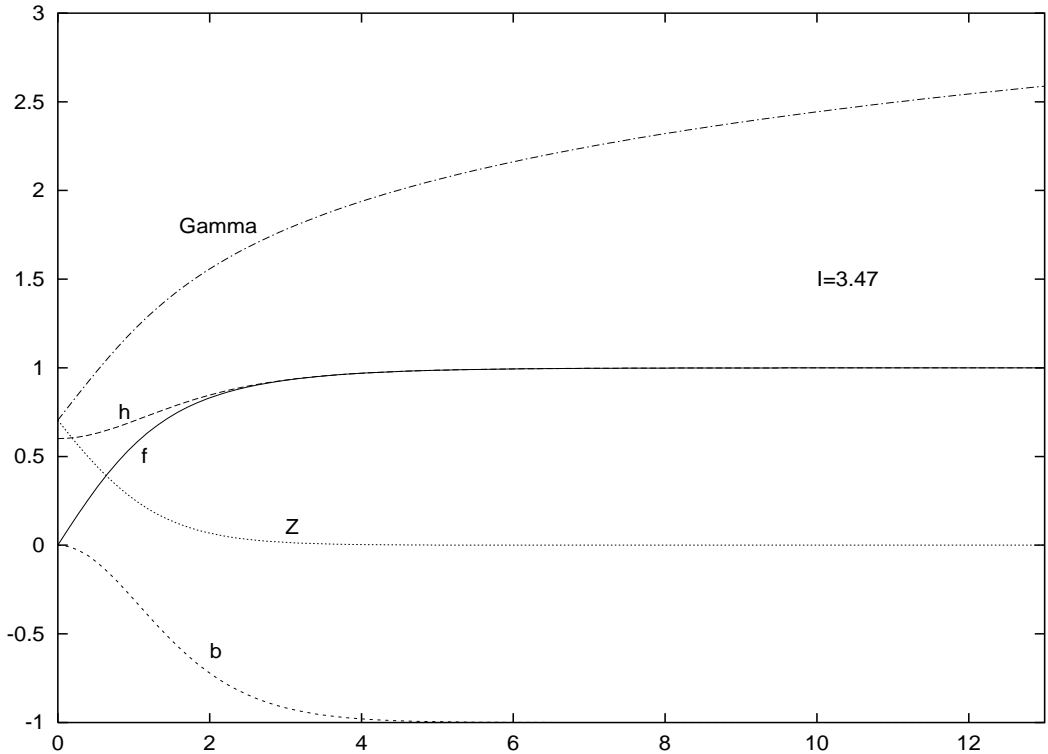} }}
\end{center}
\caption{The solutions to the full, coupled field equations for the
case of a pure current ($\epsilon=+1$) with $\lambda=0.5$,
$\kappa= 0.3$, $g=e=1$ and at infinity, $\gamma_z \rightarrow (I/2\pi)
\ln r$. Note how the component $h$ has been reduced in the vicinity of
the string core, relative to the bare string solution.} 
\label{fig:allfields}
\end{figure}

\begin{figure}
\begin{center}
    \leavevmode
        {\hbox %
{\epsfxsize = 15.8cm\epsfysize=20cm
    \epsffile {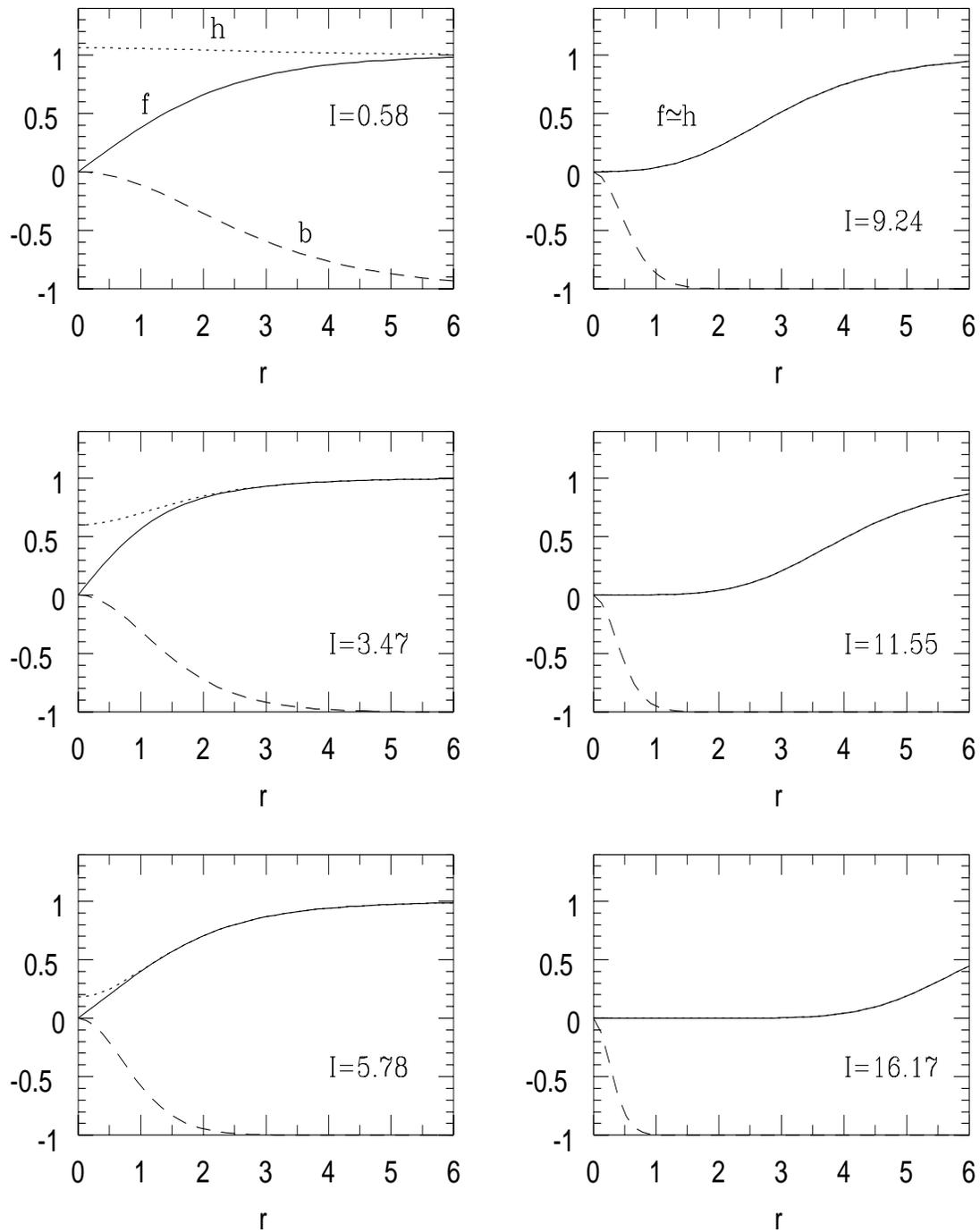} }}
\end{center}
\caption{The backreaction of a non-chiral current ($\epsilon=+1$) on
the string for progressively larger space-like currents. The couplings
are as in the previous figure.}
\label{fig:backreaction}
\end{figure}

We have performed a full numerical study of the equations of motion. A
relaxation method was employed to solve the five coupled second-order field 
equations (\ref{eqn:curr_EOM}) and (\ref{eqn:str_EOM}),
and the results are shown for $\epsilon = +1$ -- an uncharged string
-- in figures \ref{fig:allfields} and \ref{fig:backreaction}. In
figure \ref{fig:backreaction} we can
see that as the current is increased, 
the string width (given by the characteristic scales of the fields $f$
and $b$) initially decreases. At higher values of the current, the
gauge field width of the string continues to decrease and the fields
$f$ and $h$ are forced to be approximately equal, as in the Abelian
configuration, and vanish out to larger and
larger distances, increasing the energy stored in the string core and
restoring the symmetry completely near $r=0$.
The case  of a charged string with no space-like current, $\epsilon
=-1$, is more difficult to solve numerically due to the oscillatory
behaviour of the fields at large $r$, but the qualitative
features for small charge densities are that the field $h$ is increased in
value near the origin and the gauge field width of the string increases.

The studies above refer to straight strings. The case of a loop is
a more difficult to interpret.
The chiral case $\epsilon=0$, in which the electromagnetic 4-current
vanishes, is free of backreaction, as we have shown
above. This feature is common to all models of string conductivity. In
the conclusion we discuss the possibility that the dynamically stable
chiral currents may be the only stable persistent currents on
non-Abelian string loops.

\section{Strings in SO(10)}
\label{section:SO_10}
Having explored the support of a current in a simple model
of a non-Abelian string, we now consider strings formed in a realistic
Grand-Unified symmetry breaking phase transition. We hope to establish
that the features we have seen above can be embedded in such a
theory. We take the model GUT of Spin(10),
the simply-connected covering group of SO(10) and the simplest
Grand-Unified group that supports topologically stable strings. The
group can break to the Standard Model in many ways and these have been
studied and categorised according to the types of defect they produce
\cite{Rachel}. However, the pattern of symmetry breaking of most
interest to us, as it is similar to that in the SU(2) model discussed
above, is \cite{KLS}
\begin{eqnarray}
{\rm SO(10)} & \stackrel{{\bf 126}}{\longrightarrow} & {\rm SU(5)}\times
Z_2 \nonumber \\
& \stackrel{{\bf 45}}{\longrightarrow} & {\rm SU(3)_c \times SU(2)_L
\times U(1)_Y \times Z_2} \nonumber \\
& \stackrel{{\bf 10}}{\longrightarrow} & {\rm SU(3)_c \times U(1)_{em}
\times Z_2.}
\end{eqnarray}
Strings are formed at the first phase transition, with a Higgs in the
complex 126 - dimensional representation of SO(10). We embed SU(5) in
SO(10) by treating 5 - dimensional complex vectors in the fundamental
representation of SU(5) as 10 - dimensional real vectors in SO(10). The
discrete ${\rm Z_2}$ symmetry corresponds to the element $\{-1\}$ that
is not an element of SU(5). The homotopy group $\pi_1[{\rm
Spin(10)/SU(5)\times Z_2}]={\rm Z_2}$ and so stable strings are
formed. The unbroken subgroup is not simply connected, and so
a closed path in configuration space enclosing the string
corresponds to movement between the disconnected pieces of the
unbroken subgroup $H$ on the group manifold of $G$, which cannot be
continuously deformed to a point.

SO(10) has 45 Hermitian generators that are defined by the relation
\[ M_{ab} = -\frac{i}{4}[\Gamma_a, \Gamma_b] \;\;\;\;\; a,b = 1
... 10,\]
where the $\Gamma_a$ are as usual constructed iteratively from the
Pauli matrices and are $32\times32$ matrices in this
representation:
\be
\Gamma_{2k+1} = \times_{i=1}^{5} a_i, \qquad \Gamma_{2k}=
\times_{i=1}^5 b_i ,
\ee
where
\be
a_i = \left\{ 
\begin{array}{ccr}
 & I & i < k \\ & \sigma_1 & i=k \\ & \sigma_3 & i>k   
\end{array}
\right.
\qquad
b_i = \left\{ 
\begin{array}{ccr}
 & I & i < k \\ & \sigma_2 & i=k \\ & \sigma_3 & i>k   
\end{array}
\right.
\ee
We can decompose the group space of G (Spin(10)) into
five $2\times 2$ subspaces. Specifically, the five diagonal generators
of SO(10) are
\begin{eqnarray}
M_{12} & = & \half \; \sigma_3 \times I \times I \times I \times I,
\nonumber \\
M_{34} & = & \half \; I \times \sigma_3 \times I \times I \times I,
\nonumber \\
& \vdots & \nonumber \\
M_{9\,10} & = & \half \; I \times I \times I \times I \times
\sigma_3. 
\label{eqn:diag}
\end{eqnarray}
The basis we employ is conveniently expressed
in terms of linear combinations of the $M_{ab}$.
We can now write down the explicit forms of the unbroken generators
--- for example, we have the 8 generators of the SU(3) colour
symmetry, of which
\[ T_3^{{\rm SU(3)}}=\half (M_{12}-M_{34}), \]
with the corresponding $\sigma_1$ and $\sigma_2$ equivalents
\begin{eqnarray}
T^I_{12} & = & \half (M_{13}+M_{24}), \nonumber \\
T^R_{12} & = & \half (M_{14}-M_{23}) \nonumber
\end{eqnarray}
respectively; we have
\[ T_3^{{\rm SU(2)_L}}=\half (M_{78} - M_{9\, 10}) \equiv L_3\]
generating the third component of left-handed weak isospin (coupling
to the electron and antineutrino), and
\[ Y=\frac{2}{3} (M_{12} + M_{34} + M_{56}) - (M_{78} + M_{9\, 10}) \]
\[
Q= \frac{1}{3} (M_{12} + M_{34} + M_{56}) - M_{9\,10}.
\]
being the hypercharge and the electric charge generators.

States in this {\bf 32} representation can be labelled according
to the eigenvalues $\frac{1}{2} \epsilon_i$ of the diagonal operators
$M_{k, k+1}$ (see Eq.(\ref{eqn:diag})), where $\epsilon_i=\pm 1$ --
\be
\bphi_{32} \equiv | \epsilon_1 \epsilon_2 \epsilon_3 \epsilon_4
\epsilon_5\rangle. 
\ee
As it is well known, this representaion is reducible to two
{\bf 16} representations, which are characterised by the value
of the chirality operator $\chi$:
\be
\chi= -i \Gamma_1 .... \Gamma_{10}.
\ee
The sixteen left-handed fermions are then assigned to the
16-dimensional representation of positive chirality.

As we mentioned above, strings are formed when the Higgs field
associated with the GUT phase transition is in the {\bf 126}
representation, which is included in the the symmetric product
of two {\bf 16} representations
\be
{\bf 16} \times {\bf 16}= {\bf 10}_S + {\bf 120}_A + {\bf 126}_S.
\ee

After the GUT phase transition, the Higgs can be defined such that it
acquires an expectation value in the (16,16) position,
\be
|- - - - - \rangle |- - - - -\rangle  \equiv |16\rangle |16\rangle,
\ee
which is a singlet under SU(5) gauge transformations.

The criterion for the strings formed to be topologically stable is
that the string generator $T_s$ satisfies the relation $e^{2\pi i T_s}
= -U$, where $ U \in$ SU(5). The simplest string 
solution is generated by the diagonal generator 
\be
T^U =\frac{1}{5} (M_{12} +M_{34}+M_{56}+M_{78}+M_{9\,10})
\ee 
that commutes with all the elements of SU(5), including of course the
electromagnetic generator. This is the most
symmetric way of generating a U(1) subgroup within SO(10) that contains
the $Z_2$ symmetry, and the Higgs and gauge fields for this
type of string are of the usual Nielsen-Olesen form. However, Aryal
and Everett \cite{AE} and Ma \cite{Ma} examined this solution and
found it to be of a higher energy per unit length than that generated
by 
\be
T_s \equiv  \half (M_{59}-M_{6\,10})=
\frac{1}{4}(I\times I\times \sigma_1 \times \sigma_3
\times \sigma_2 + I\times I\times \sigma_2 \times \sigma_3
\times \sigma_1),
\ee
a linear combination of the generators of right-handed weak
isospin. At the level of this first symmetry breaking, there is a
degeneracy of possible choices for $T_s$, but this degeneracy is
progressively lifted at subsequent breakings by a combination of
energetic considerations and demanding that the generators
corresponding to physical observables be single-valued \cite{GoldBuch}.
This generator does {\it not} commute with charge.

Together with $T_s$, the two generators
\ba
A =  \half (M_{56} + M_{9\, 10}) &=&\frac{1}{4}(I \times I \times I
\times I \times \sigma_3 \pl 
I \times I \times  \sigma_3 \times I \times I), \\ 
B = \half (M_{5\, 10} + M_{69}) & = & \frac{1}{4}(I \times I \times \sigma_2
\times I \times \sigma_2 \mi I \times I \times
\sigma_1 \times I \times \sigma_1).
\ea
generate an SU(2) algebra,
\[
[T_s,A]=-iB,\qquad [T_s,B]=iA, \qquad [B,A]=iT_s.
\]
We introduce the generator
\ba
T_0 &=&-\frac{1}{2}(M_{12} + M_{34})  - M_{56} + M_{9\, 10} \\
 & = &-\frac{1}{4} (\sigma_3 \times I  \times I \times I \times I
\pl  I \times \sigma_3 \times I \times I \times I\\
& & \pl  2 I \times I\times
\sigma_3 \times I \times I \mi 2 I \times I \times I \times I \times
\sigma_3),
\ea
which commutes with $T_s$, and so the charge generator breaks into the
components 
\[
Q= -\frac{2}{3}(A + T_0).
\]
The subspace of the {\bf 126}
representation in which these generators act is 
generated by the states
\be
\left\{ |14\rangle |14\rangle, \;\;\;
 \frac{1}{\sqrt{2}} \left( |14\rangle|16\rangle +
|16\rangle|14\rangle \right),
\;\;\; |16\rangle|16\rangle \right \},
\label{eqn:subs}
\ee
where
\be
|14\rangle=|- - + - +\rangle.
\ee
The form of the generators in this subspace is then
\[
T_s={1\over\surd2}\pmatrix{0&1&0\cr1&0&1\cr0&1&0},\qquad
A=\pmatrix{1&0&0\cr0&0&0\cr0&0&-1},\qquad
B={1\over\surd2}\pmatrix{0&-i&0\cr i&0&-i\cr0&i&0},
\]
while $T_0$ is the identity matrix. It is straightforward to check
that these generators are related to the ones we discussed in
section \ref{section:simple_z2_string} by the
transformations
\be
A=C^{-1} T_2 C, \qquad  B= C^{-1} T_1 C,  \qquad T_s=C^{-1} T_3 C,
\label{eqn:corres}
\ee
where
\[
C={1\over\surd2} \pmatrix{1&0&1\cr0&i\surd2&0\cr-i&0&i}.
\]
After this change of base then, the model becomes identical
to the SU(2)$\times$U(1)  model above.
For instance, the covariant derivative
\[
D_{\mu} \Phi = (\partial_{\mu} + ig (A_{\mu}^{A} A + A_{\mu}^B B +A_{\mu}^T T
+ A_{\mu}^0 T_0 ))\Phi
\]
is equivalent to the SU(2)$\times$U(1) definition in 
Eqs.(\ref{eqn:definitions}) after the obvious identification of gauge fields
as indicated in Eq.(\ref{eqn:corres}) and setting the two gauge
couplings $e$ and $g$ to be equal.

For the Abelian ansatz, notice that in the subspace defined in
equation \ref{eqn:subs}, the generator $T^U$ is given by
\be
T^U= \frac{1}{5} ( 2A - 3 I),
\ee
and after the change of basis (equation \ref{eqn:corres}),
\be
T^U= \frac{1}{5} ( 2T_2 -3I).
\ee
In this ansatz, then, the covariant derivative
of the Higgs field reads
\be
D_\mu\Phi= (\partial_\mu + i\frac{g}{5} (-3I + 2 A)A_\mu)) \Phi.
\ee
In the SU(2)$\times$U(1) model, we had
\[
D_\mu \Phi = (\partial_\mu - i(-eA_\mu + gB_\mu^2 T_2)) \Phi. \\
\]
We define
\[
A_\mu = \frac{e}{\alpha} Z_\mu, \qquad B^2_\mu= -\frac{g}{\alpha}
Z_\mu,
\]
where $\alpha=\sqrt{e^2 + g^2}$. We can see that in this case,
$e^2=(3/2)g^2$.

\section{Conclusion}
\label{section:conclusion}

We have seen that the essential feature of non-Abelian string
conductivity models is that the string generator does not commute with
electromagnetic charge. If it did, the Higgs would be annihilated by
the charge generator at all distances from the string core, and so
would be uncharged. It is precisely because $T_s$ and $Q$ do not
share a common set of eigenvectors, and that different eigenvectors of
$T_s$ have different profiles (in the discussion in section
\ref{section:simple_z2_string}, $f$ and $h$) near the string core,
that a charged component of the Higgs exists there. 

Charge is, however, generically ill-defined at the origin. 
The charged component of the Higgs (see equation \ref{eqn:base_decomp})
cannot really be considered to be a well-defined `condensate' in the
sense of the Witten model. Moreover, 
we cannot universally give this charged   
component a varying phase with an electromagnetic gauge
transformation around a loop of such string, as it does not
vanish at $r=0$ where $Q$ itself is ill-defined.
If we were to try to construct such a solution we would have to give
the Higgs a phase with a generator that commuted with $T_s$ at $r=0$ and
deformed into $Q$ outside the string. That implies there
must be some radial dependence of the Lie algebra element that acts on
the condensate, 
and we cannot construct a non-trivial $r$ and $z$-dependent
condensate that is periodic in $z$, as is required for a loop.  This
can be seen as follows. Following the infinite-string ansatz of Alford
{\it et al.} \cite{Alford} for a zero-mode excitation,  we require a
Higgs of the form 
\[
\Phi(r,\theta, z,t) = e^{i\beta(z,t){\bf
S}(r,\theta)}\bar{\Phi}(r,\theta), 
\]
where $\bar{\Phi}$ is the unperturbed string solution, and ${\bf S}$ lies in
the Lie alebra of the unbroken group $H$ and tends 
to charge $\bf{Q}(\theta)$ away from the string core.
We wish to implement this on the topology of a loop of length $L$.
Now the coordinate $z$ measures the distance along the string.
The requirement
\[
\Phi(z+L) = \Phi(z)
\]
implies that
\be
\beta(z+L,t){\bf S}(r,\theta) \equiv \beta(z,t){\bf S}(r,\theta) +
2\pi N 
\mbox{\boldmath $1$}.
\ee
This cannot be satisfied everywhere  unless $\beta$ is
trivial, or $S$ is independent of $r$ and $\theta$ (which is not
possible for the case $[T_s,Q] \neq 0$ we require).
We see, then, that we must view current excitations on
such non-Abelian strings as
being supported by the string gauge field, which is also charged, and
sources excitations of the form
\be
{\bf A}_\alpha = -{\bf S}(r,\theta) \beta (z,t).
\ee
This viewpoint is consistent with the topology of a string loop.

The boundary conditions on the gauge field equations of motion
(which, in the absence of a Higgs phase, are homogeneous in $S$)  now
demand that the ${\bf A}_\alpha$ vanish both at the center of the
string loop and at spatial infinity \cite{Landau}. The lowest energy
solution in the 
absence of an external field is simply ${\bf S}\equiv 0$ and hence we
suggest that an uncharged non-Abelian string loop ($\epsilon > 0$)
will be incapable of 
supporting a persistent current in this case, making it in effect a
perfect conductor. We aim to explore the issue of perfect conductor
electrodynamics in a subsequent paper\footnote{Work in
progress.}. Similarly, we 
would not expect the charged $\epsilon < 0$ 
case to persist either. It is conceivable that the chiral state, in
which the current is identified with the charge density, {\it
could} be dynamically stable, however. The solution of the equations
of motion on a string loop need to be solved fully to confirm this,
however, to take into account curvature effects.

Studies of current-carrying loop stability have to date worked with
the Witten model of a superconducting string, in which a loop (or {\it
vorton}) has two
associated conserved quantities $N$ and $Z$; these correspond to the
condensate winding and the net charge carried by the loop. In terms of the
function $\beta(z,t)=kz \pm \omega t$ defined on the string
worldsheet, these come from the first term (a topological invariant)
and second term (Noether charge)  respectively. The fact that
there is no condensate in non-Abelian models seems to preclude the existence
of the quantum number $N$. However, it may be that the {\it chiral}
current-carrying states may be dynamically
stable on a loop, with a net charge $Q$. These may be
vulnerable to the loss 
of charge carriers at cusps \cite{Cop}, however, and so
these currents can leak from the string when the loop has a contorted
small-scale structure. 

Despite the questions raised about the stability of currents, it seems
that there are so many competing perfect-conduction mechanisms
clamouring for a string's attention that it seems almost certain that
if strings exist at all they will be current-carrying at some
scale. It seems that if GUT strings are formed they are
likely to be capable of carrying enormous currents -- we see that the
Euler-Lagrange equations provide no natural upper limit, and
so quantum effects must intervene through particle production in the
vicinity of the string.
Chiral current magnitudes may be limited 
by the amount of charge a loop can randomly
acquire, either at formation or after intercommutation of strings
carrying different currents. If the bare GUT string does not support currents,
whether fermionic or vector bosonic, then 
subsequent dressing at the electroweak scale appears to provide
them. This provides a further degree of freedom for string-based
models of structure formation and dark matter.

Note that the {\it induced} vector-boson
conductivity models \cite{AnneWarren1}-\cite{AnneWarren2} are not
included in this. They are complementary to the models listed and will
presumably display similar topological currents and/or zero modes
according to whether charge commutes with the string generator or not.

In this paper we have explored non-Abelian string conductivity in some
detail. We have emphasised the
role of topology in true superconductivity, and its absence in
non-Abelian string models. We have analysed
the current solution (with and without backreaction) in an
illustrative non-Abelian string model. We have also shown that the
standard, lowest-energy 
SO(10) string solution is a simple embedding of this model and hence
also supports currents.
We suggest that only charged non-Abelian strings will be able to
carry persistent currents in the absence of an external field, due to
dynamical stability of the chiral zero-modes. It seems that vorton
stability is less likely in realistic GUT string models, then, which avoids
the overdensity problem and reinforces structure formation scenarios
involving defects at such a scale.
\section*{Acknowledgments}
G.L. thanks the Alexander von Humboldt Foundation for
financial support at Heidelberg University. G.L also  
thanks Fundaci\'on Antorchas
and The British Council for partial financial support
at Imperial College, where this work was initiated.
This work was supported in part
by the European Commission under the Human Capital and Mobility
programme, contract no. CHRX-CT94-0423.


\begin{thebibliography}{99}

\bibitem{Witten} E. Witten, {\it Nucl. Phys. B} {\bf 249}, 557 (1985)
\bibitem{Ostriker} J.P. Ostriker, C. Thompson, E. Witten, {\it
Phys. Lett. B} {\bf 180}, 231 (1986)
\bibitem{Levin} J.J. Levin, K. Freese, D.N. Spergel, {\it
Astrophys. J.} {\bf 389}, 464 (1992)
\bibitem{Hill} See, for example,
C.T. Hill, D. Schramm, T. Walker,  {\it Phys. Rev. D} {\bf
36}, 1007 (1987)
\bibitem{Chud} E.M. Chudnovsky, G.B. Field, D.N. Spergel, A. Vilenkin,
{\it Phys. Rev. D} {\bf 34}, 944 (1986)
\bibitem{vortons} See, for instance, B. Carter, `The Mechanics of
Cosmic Rings', {\it Phys. Lett. B} {\bf 238}, 166 (1990); B. Carter,
X. Martin, {\it Ann. Phys.} {\bf 227}, 151 (1993);  B. Carter,
P. Peter, A. Gangui, preprint {\tt hep-ph/9609401} (1996)
\bibitem{Everyone} R. Brandenberger, B. Carter, A.-C. Davis,
M. Trodden, {\it Phys. Rev. D} {\bf 54}, Vol. 10 (1996) 
\bibitem{reviews} A. Vilenkin, E.P.S. Shellard, {\it Cosmic Strings and other
Topological Defects}, Cambridge University Press (1994), 
M.B. Hindsmarsh, T.W.B. Kibble, {\it Rep. Prog. Phys} {\bf 58},
477 (1995).
\bibitem{Presk} J. Preskill,  Proceedings of Les Houches Summer
School, {\it Architecture of Fundamental Interactions at Short 
Distances}, Jul 1 - Aug 8, 1985.  Eds. Pierre Ramond, Raymond
Stora. North-Holland (1987)
\bibitem{Everett} A.E. Everett, {\it Phys. Rev. Lett.} {\bf 61}
16, 1807 (1988)
\bibitem{Alford} M. Alford, K. Benson, S. Coleman, F. Wilczeck,
 J. March-Russell, {\it Nucl. Phys. B} {\bf 349}, 414 (1991)
\bibitem{Amb} J. Ambj\o rn, P. Olesen, {\it Int. Jour. Mod. Phys. A}
{\bf 5} No. 23, 4525 (1990)
\bibitem{AnneWarren1} A.C. Davis,  W. Perkins,
{\it Nucl. Phys. B} {\bf 406}, 377 (1993)
\bibitem{Warren} W. Perkins, {\it Phys. Rev. D} {\bf 47}, R5224 (1993)
\bibitem{AnnePat} A.C. Davis, P. Peter, {\it  Phys. Lett. B} {\bf 358},
197 (1995)
\bibitem{AnneWarren2} A.C. Davis,  W. Perkins, preprint {\tt
hep-ph/9610292} (1996) 
\bibitem{DavisShell} R.L. Davis, E.P.S. Shellard, {\it  Phys. Lett. B}
{\bf 207}, 404 {\it and} {\bf 209}, 485 (1988)
\bibitem{deVega} H.J. de Vega, F.A. Schaposnik, {\it Phys. Rev. Lett.}
{\bf 56},2564 (1986)
\bibitem{Khare} C.N. Kumar, A.Khare, {\it Phys.Lett. B} {\bf 178} 178, 395 (1986);{\it Phys. Rev. D} { \bf 36}, 3253 (1987)
\bibitem{Bogo} E.B. Bogomol'nyi, {\it Sov. J. Nucl. Phys.}
{\bf 24} No. 4, 449 (1976); H.J. de Vega, F.A. Schaposnik, {\it Phys.Rev. D}
{\bf 14}, 1100, (1976); for similar equations in the SU(2) model,
see L. Cugliando, G. Lozano, F.A. Schaposnik, {\it Phys. Rev. D} {\bf
40}, 3440 (1989)
\bibitem{Ma} C.P. Ma,  {\it Phys. Rev. D} {\bf 48}, 530 (1993)
\bibitem{AE} M. Aryal, A.E. Everett, {\it Phys. Rev. D} {\bf 35}, 3105
(1987)
\bibitem{Rachel} R. Jeannerot, A.-C. Davis, {\it Phys. Rev. D} {\bf
52}, 7220 (1995)
\bibitem{KLS} T.W.B. Kibble, G. Lazarides, Q. Shafi, {\it
Phys. Lett. B} {\bf 113}, 237 (1982)
\bibitem{GoldBuch} M. Bucher, A.S. Goldhaber, {\it Phys. Rev. D}
{\bf 49}, 4167 (1994)
\bibitem{Landau} L. Landau and L. Lifshitz, `The Electrodynamics of
Continuous Media', Pergamon Press, Oxford (1984)
\bibitem{Davis88} R.L. Davis, {\it Phys. Rev. D} {\bf 38}, 3722 (1988)
\bibitem{Cop} E.J. Copeland, D. Haws, M.B. Hindmarsh,
N. Turok, {\it Nucl. Phys. B} {\bf 306}, 908 (1988)
\end{thebibliography}
\end{document}